\documentstyle[12pt]{article}
\def\ZZZ{{\hbox{ Z\kern-1.6mm Z}}}
\def\zzz{{\hbox{ z\kern-1mm z}}}

\newcommand{\EE}{{\cal E}}

\newcommand{\LL}{{\cal L}}

\newcommand{\wh}{\widehat}

\newcommand{\RR}{{\cal R}}

\newcommand{\be}{\begin{equation}}
\newcommand{\ee}{\end{equation}}
\newcommand{\ben}{\begin{eqnarray}\displaystyle}
\newcommand{\een}{\end{eqnarray}}
\newcommand{\refb}[1]{(\ref{#1})}
\newcommand{\p}{\partial}
\newcommand{\sectiono}[1]{\section{#1}\setcounter{equation}{0}}

\def\one{{\hbox{ 1\kern-.8mm l}}}
\def\zero{{\hbox{ 0\kern-1.5mm 0}}}

\begin{document}

{}~
{}~

\hfill\vbox{\hbox{hep-th/0601228}}\break

\vskip .6cm

\medskip

\baselineskip 20pt 

\begin{center}

{\Large \bf
 BTZ Black Hole
with  Chern-Simons and Higher Derivative Terms}

\end{center}

\vskip .6cm
\medskip

\vspace*{4.0ex}

\centerline{\large \rm Bindusar Sahoo and Ashoke Sen}

\vspace*{4.0ex}

\centerline{\large \it Harish-Chandra Research Institute}

\centerline{\large \it  Chhatnag Road, Jhusi,
Allahabad 211019, INDIA}

\vspace*{1.0ex}

\centerline{E-mail: bindusar@mri.ernet.in, ashoke.sen@cern.ch,
sen@mri.ernet.in}

\vspace*{5.0ex}

\centerline{\bf Abstract} \bigskip

The entropy of a BTZ black hole in the 
presence of  gravitational Chern-Simons terms has previously 
been analyzed using
Euclidean action formalism. In this paper we 
treat
the BTZ solution
as a two dimensional black hole by regarding the angular 
coordinate  as a compact direction, and 
use
Wald's Noether charge method to
calculate the entropy of this
black hole in the 
presence of higher derivative and 
gravitational Chern-Simons terms.  
The parameters labelling the black hole solution can be 
determined
by extremizing an entropy function whose value at the extremum
gives the entropy of the black hole.

\vfill \eject

\baselineskip 18pt

\tableofcontents

\sectiono{Introduction} \label{s1}

BTZ solution describes a black hole in three 
dimensional theory of gravity with negative cosmological 
constant\cite{9204099} and often
appears as a factor in the near horizon geometry of
higher dimensional black holes in string theory\cite{9712251}. 
For this
reason
it has provided us with a useful 
tool for relating black hole entropy to the degeneracy of 
microstates of the black hole, both in three dimensional
theories of gravity and also in string theory\cite{9712251,brown}. 
Initial studies involved relating the Bekenstein-Hawking formula
for BTZ black hole entropy in two derivative theories of gravity
to the Cardy formula
for the degeneracy of states 
in the two dimensional conformal field theory 
living on the asymptotic boundary. 
Later this was 
generalized to higher derivative theories of gravity, where the lagrangian 
density contains arbitrary powers of Riemann tensor and its covariant 
derivatives\cite{9909061}. For computing entropy of black holes in such 
theories one can no longer use the area formula. Instead one needs to
use 
the Noether charge method developed by 
Wald\cite{9307038,9312023,9403028,9502009}.

In three dimensions one can also
add to the action the gravitational 
Chern-Simons terms. In this case the 
Lagrangian density cannot be written in a 
manifestly covariant form and as a result
Wald's formalism cannot be applied 
in a straightforward fashion. For this reason
the effect of this term on the black 
hole entropy was analyzed in \cite{0506176,0508218}
using the Euclidean action formalism\cite{9804085}. A different
Euclidean method yielding the same result can be found in
\cite{0509148}.

The 
goal of this paper is to compute the entropy of BTZ
black holes in the presence of Chern-Simons and higher derivative
terms using Wald's Noether charge method.
In order to do this we regard the BTZ black hole as a two dimensional 
configuration by treating the angular coordinate as a 
compact direction\cite{9304068}. 
The black hole entropy is then calculated using the dimensionally 
reduced two dimensional theory.  This has the advantage that the 
Chern-Simons term, which was not manifestly covariant 
in three dimensions, 
reduces to a manifestly covariant set of terms in two 
dimensions\cite{0305117}. 
Hence Wald's formula for the black hole entropy can be applied in a 
straightforward fashion. The result agrees with the one calculated using
the Euclidean action formalism.

The rest of the paper is organized as follows. In section \ref{s2} we 
discuss the dimensional reduction of a general three dimensional theory of 
gravity, including the gravitational Chern-Simons term, to two dimensions 
and describe the BTZ solution from two dimensional viewpoint. In section 
\ref{s3} we calculate the entropy of extremal BTZ black holes using the 
entropy 
function formalism\cite{0506177} which is known to be equivalent to Wald's 
Noether charge method. In section \ref{s4} we calculate the 
entropy of a non-extremal BTZ black hole using Wald's method directly. 
Both for the extremal and the non-extremal black holes the parameters 
labelling the solution can be obtained by extremizing an 
entropy function whose value at the extremum gives the entropy.

\sectiono{The Two Dimensional View} \label{s2}

Let us consider a three dimensional theory of gravity with
metric $G_{MN}$ ($0\le M,N\le 2$) and a general 
action of the form:\footnote{We could
add any number of scalar
fields without changing the final result since they must be frozen
to constant values in order to comply with the homogeneity of the
BTZ configuration.}
\be \label{e1}
S = \int d^3 x \sqrt{-\det G}\left[ \LL^{(3)}_0
+ \LL^{(3)}_1 \right]\, .
\ee
Here
$\LL^{(3)}_0$ denotes an arbitrary scalar constructed out of the metric, 
the 
Riemann tensor and covariant derivatives of the Riemann tensor. On
the other hand $\sqrt{-\det G}\, \LL^{(3)}_1$ denotes 
the gravitational Chern-Simons term:
\be \label{e1a}
\sqrt{-\det G}\, \LL^{(3)}_1 = K \, \Omega_3(\wh\Gamma) \, ,
\ee
where
$K$ is a 
constant, $\wh\Gamma$ is the Christoffel connection constructed out of 
the metric $G_{MN}$ and
\be \label{e2}
\Omega_3(\wh\Gamma) = \epsilon^{MNP} \left[{1\over 2}
\wh\Gamma^R_{MS} \p_N \wh\Gamma^S_{PR} + {1\over 3}
\wh\Gamma^R_{MS} \wh\Gamma^S_{NT} 
\wh\Gamma^T_{PR}\right]\, .
\ee
$\epsilon$ is the
totally anti-symmetric symbol with $\epsilon^{012}=1$.

We shall consider field configurations where one of the coordinates (say 
$y\equiv x^2$) is compact with 
period $2\pi$ and the metric is independent 
of this compact direction. In this case we can define two 
dimensional fields through the relation:
\be \label{e3}
G_{MN} dx^M dx^N = \phi \left[ g_{\mu\nu} 
dx^\mu dx^\nu + (dy + A_\mu 
dx^\mu)^2\right]\, .
\ee
Here $g_{\mu\nu}$ ($0\le \mu,\nu\le 1$) denotes a 
two dimensional metric, $A_\mu$ denotes a two dimensional 
gauge  field and $\phi$ denotes a 
two dimensional scalar field. In terms of these 
two dimensional fields the 
action takes the form:
\be \label{e4}
S = \int d^2 x \sqrt{-\det g}\left[ \LL^{(2)}_0 + \LL^{(2)}_1 \right]
\ee
where
\be \label{e5}
\sqrt{-\det g} \, \LL^{(2)}_0 = \int dy \sqrt{-\det G} \,
\LL^{(3)}_0
= 2\pi \sqrt{-\det G} \, \LL^{(3)}_0\, ,
\ee
and\cite{0305117}
\be \label{e5a}
\sqrt{-\det g} \, \LL^{(2)}_1 = K \, \pi \, \left[
 {1\over 2} R \varepsilon^{\mu\nu} F_{\mu\nu}
+{1\over 2} \varepsilon^{\mu\nu} F_{\mu\tau} F^{\tau\sigma} 
F_{\sigma\nu} \right]\, .
\ee
Here
$R$ is the scalar curvature of the two dimensional metric 
$g_{\mu\nu}$:
\ben \label{e7}
&& \Gamma^\mu_{\nu\rho} = {1\over 2} g^{\mu\sigma}
\left( \p_\nu g_{\sigma\rho} + \p_\rho g_{\sigma\nu}
- \p_\sigma g_{\nu\rho} \right) \nonumber \\
&& R^\mu_{~\nu\rho\sigma} = \p_\rho
\Gamma^\mu_{\nu\sigma} - \p_\sigma
\Gamma^\mu_{\nu\rho} + \Gamma^\mu_{\tau\rho}
\Gamma^\tau_{\nu\sigma} - \Gamma^\mu_{\tau\sigma}
\Gamma^\tau_{\nu\rho} \nonumber \\
&& R_{\nu\sigma} = R^\mu_{~\nu\mu\sigma}, \qquad
R = g^{\nu\sigma}  R_{\nu\sigma}\, ,
\een
$\varepsilon^{\mu\nu}$ is the totally antisymmetric symbol with
$\varepsilon^{01}=1$, and
\be \label{e6}
F_{\mu\nu} = \p_\mu A_\nu - \p_\nu A_\mu\, .
\ee
\refb{e5} follows in a 
straightforward 
fashion from \refb{e1}.
\refb{e5a}
comes from dimensional reduction of the Chern-Simons term after
throwing away total derivative terms and was 
worked out in 
\cite{0305117}. Note that although the 
Chern-Simons term cannot be expressed in a 
manifestly covariant form in three dimensions, it does reduce to a 
manifestly covariant expression in two dimensions.

A general BTZ black hole in the three dimensional theory
is described by the metric:
\be \label{e16}
G_{MN} dx^M dx^N = -{
(\rho^2 - \rho_+^2) (\rho^2 - \rho_-^2)\over l^2 \rho^2} d\tau^2
+ {l^2 \rho^2 \over (\rho^2 - \rho_+^2) (\rho^2 - \rho_-^2)} d\rho^2
+ \rho^2 \left(dy - {\rho_+ \rho_-\over l \rho^2} d\tau\right)^2\, ,
\ee
where $l$, $\rho_+$ and $\rho_-$ are parameters labelling the
solution.\footnote{Since the local geometry of a BTZ black hole is 
that of $AdS_3$ which is maximally symmetric, the higher derivative 
corrections do not change the structure of the solution. However for a 
black hole carrying a given mass and angular momentum, the values of the 
parameters $l$, $\rho_+$ and $\rho_-$ depend on the higher derivative 
terms.}
Comparing this with \refb{e3} gives\cite{9304068}
\ben \label{e17}
&& \phi = \rho^2\, , \qquad
A_\mu dx^\mu = - {\rho_+ \rho_-\over l \rho^2} d\tau, \nonumber \\
&& g_{\mu\nu} dx^\mu dx^\nu 
= -{
(\rho^2 - \rho_+^2) (\rho^2 - \rho_-^2)\over l^2 \rho^4} d\tau^2
+ {l^2 \over (\rho^2 - \rho_+^2) (\rho^2 - \rho_-^2)} d\rho^2
\, .
\een

\sectiono{Extremal BTZ Black Holes} \label{s3}

We shall define
a general extremal black hole in the two dimensional theory 
to 
be the one whose near horizon geometry is $AdS_2$ and for 
which the scalar 
field $\phi$ and 
the gauge field strength $F_{\mu\nu}$ are invariant under 
the $SO(2,1)$ isometry of the $AdS_2$ background\cite{0506177}. 
The most general near 
horizon background consistent with this requirement is
\be \label{e8}
g_{\mu\nu} dx^\mu dx^\nu = v\left(-r^2 dt^2 + {dr^2\over r^2}\right), 
\qquad F_{rt} = e, \qquad \phi = u\, ,
\ee
where $v$, $e$ and $u$ are constants. 
Following \cite{0506177,0508042} we define
\be \label{e9}
f(u, v, e) = \sqrt{-\det g}\,  (\LL^{(2)}_0+\LL^{(2)}_1)\, ,
\ee
evaluated in the background \refb{e8}, and
\be \label{e10}
\EE(u, v, e, q) = 2\pi (eq - f(u, v, e))\, .
\ee
As was shown in \cite{0506177}, 
the near horizon values of $u$, $v$ and $e$ for an extremal 
black hole with electric charge $q$ is obtained by extremizing the 
`entropy 
function' $\EE$ with respect to these variables. Furthermore, Wald's entropy 
for this black hole is given by the value of the function $\EE$ at this 
extremum\cite{0506177}. 

Using eqs.\refb{e5}, \refb{e5a} and \refb{e8},
\refb{e9} we see that
for the theory considered here,
\be \label{e10a}
f(u, v, e) = f_0(u,v,e) + \pi\, K\, (2\, e \,
v^{-1} - e^3\, v^{-2})\, ,  
\ee
where
\be \label{e10b}
f_0(u, v, e) = {2\pi}\, \sqrt{-\det G}\,  \LL^{(3)}_0 \, .
\ee

Let us now specialize to the case of extremal BTZ black holes.
These correspond to choosing
$\rho_-=\pm
\rho_+$ in \refb{e16}, \refb{e17} 
and the near horizon
limit is obtained by taking $\rho$ close to $\rho_+$. Defining
\be \label{e18}
r=\rho-\rho_+, \qquad t = {4 \over l^2} \, \tau\, ,
\ee
we can express \refb{e16}
\refb{e17} for $\rho_-=\pm \rho_+$ and small $r$
as
\be \label{e3d}
G_{MN} dx^M dx^N = {l^2\over 4}\, \left( - r^2 dt^2 
+ {dr^2\over r^2}\right) + \rho_+^2 \, \left( 
dy \pm
\left(- {l\over 4}  + {l\over 2\rho_+}\, r
\right) \, dt
 \right)^2
\ee
\be \label{e19}
\phi = \rho_+^2\, , \qquad
A_\mu dx^\mu = \pm\left(- {l\over 4}  + {l\over 2\rho_+}\, r
\right) \, dt, \qquad g_{\mu\nu} dx^\mu dx^\nu =
{l^2\over 4 \rho_+^2} \, \left( - r^2 dt^2 + {dr^2\over r^2}\right)
 \, .
\ee
Comparison with \refb{e8} now yields
\be \label{e20}
u = \rho_+^2, \qquad v={l^2\over 4\rho_+^2}, \qquad
e = \pm {l\over 2 \rho_+}\, .
\ee
Note that instead of three independent parameters $u$, $v$ and $e$,
we now have two independent parameters $l$ and $\rho^+$ 
labelling the near 
horizon
geometry.  In particular  $v$ and $e$ satisfy the relation
\be \label{e20a}
v = e^2\, .
\ee
This is a reflection of the fact that the BTZ black hole is
locally $AdS_3$ and hence
has a higher degree of symmetry than the more general configuration
considered in \refb{e8}. This is a consistent truncation of the parameter
space and hence we can extremize the entropy function $\EE$
subject to this constraint. We shall  choose $e$ and 
\be \label{e21}
l = 2\sqrt{ue^2}\, ,
\ee 
as independent
variables.  

Our next step will be to investigate the structure of $f_0$ given in
\refb{e10b} for the extremal BTZ black hole 
solution described above.
Since the BTZ black hole is locally the maximally  symmetric $AdS_3$ space, $\LL^{(3)}_0$, being a scalar
constructed out of the Riemann tensor and its covariant derivatives,
must be a constant. Furthermore since locally BTZ metrics for
different values of $\rho_\pm$ are related by coordinate
transformation, $\LL^{(3)}_0$ must be independent of
$\rho_\pm$ and hence is a function of $l$ only.
Let us define 
\be \label{en2}
h(l) = \LL^{(3)}_0
\ee
evaluated in the BTZ black hole geometry.  (Note that this definition
is independent of whether we are using the extremal or non-extremal
metric.) Since for the extremal black hole metric \refb{e3d}
 \be \label{en3}
\sqrt{-\det G} = {l^2 \rho_+ \over 4} = {l^3 \over 8\, |e|}\, ,
\ee
we get
\be \label{e22}
f_0 = 2\pi \, \sqrt{-\det G} \,
\LL^{(3)}_0  =   {1\over |e|} g(l)\, ,
\ee
where 
\be \label{en6}
g(l) = {\pi \, l^3\, h(l)\over 4}\, .
\ee
 Eqs.\refb{e10}, \refb{e10a},
\refb{e20a} and \refb{e22} now give
\be \label{e23}
\EE = 2\pi \left( q\, e - {1\over |e|} g(l) - {\pi \, K\over e}
\right)\, .
\ee
We need to extremize this with respect to $l$ and $e$. The
extremization with respect to $l$ requires extremization of
$g(l)$ with respect to $l$. Let us define
\be \label{ecdef}
C = -{1\over \pi} g(l)
\ee
at the extremum of $g$. This gives
\be \label{e24}
\EE = 2\pi \left( q\, e + {\pi\, C\over |e|} - {\pi\, K\over e} 
\right) \, .
\ee
We shall assume that $C\ge |K|$. 
Extremizing \refb{e24} with respect to $e$ we now get:
\ben \label{eevalue}
e &=& \sqrt{\pi (C-K)\over q} \quad \hbox{for} \, q>0\, ,
\nonumber \\
&=& \sqrt{\pi (C+K)\over |q|} \quad \hbox{for} \, q<0\, .
\een
Furthermore,  at the extremum,
\ben \label{e25}
\EE &&= 2\pi \sqrt{ c_R \, q\over 6} \quad \hbox{for} \, q>0\, ,
\nonumber \\
&&= 2\pi \sqrt{c_L \, |q|\over 6} \quad \hbox{for} \, q<0\, ,
\een
where we have defined
\be \label{e26}
c_L = 24\, \pi\, (C+K)\, , \qquad c_R = 
24\, \pi\, (C-K) \, .
\ee
\refb{e25} gives the entropy of extremal BTZ black hole.
Since the conserved charge $q$ measures 
momemtum along $y$, which
for a BTZ black hole represents the angular 
momentum $J$, eqs.\refb{e25}, \refb{e26} are in agreement
with the results of \cite{0506176} for extremal BTZ black holes
with mass = $|J|$. 

Note that the Chern-Simons term plays no role in the determination
of the parameter $l$ and
\be \label{e26a}
c_L+c_R = 48\, \pi \, C = -48\, g(l)\, .
\ee
This is a reflection of the fact that in three dimensions the effect of 
the Chern-Simons term on the equations of motion involves covariant 
derivative of the Ricci tensor\cite{0305117} which vanishes for BTZ 
solution. On the other hand \be \label{e27}
c_L - c_R = 48\, \pi \, K \, ,
\ee
is insensitive to the detailed structure of the higher derivative
terms and is determined completely by the coefficient of the 
Chern-Simons term. In the analysis of \cite{0506176} this was
a consequence of the fact that $c_L-c_R$ is related to the
diffeomorphism anomaly of the bulk theory. In the present
context this is a consequence of the fact that $c_L-c_R$ is
determined by the parity odd part of the action evaluated
on the near horizon geometry of the BTZ black hole, 
and this contribution 
comes solely from the Chern-Simons term.

\sectiono{Non-extremal BTZ Black Holes} \label{s4}

We now turn to the computation of the entropy of a general
non-extremal BTZ black hole solution given in
eqs.\refb{e16}, \refb{e17}. 
First we note that the local
geometry of extremal and non-extremal black holes are identical since
both describe a locally $AdS_3$ space-time of curvature radius $l$.
Thus $l$ for a non-extremal black hole is determined by the same 
equation as in the extremal case, \i.e. via the extremization of the
function $g(l)$:
\be \label{en1}
g'(l)=0\, .
\ee
Since the contribution to the Noether charge from different terms
in the action add, 
the entropy computed from Wald's general 
formula\cite{9307038,9312023,9403028,9502009} can be regarded
as the sum of two terms, -- one arising from the $\LL^{(3)}_0$
term in the action and the other arising out of the $\LL^{(3)}_1$
term in the action. In the $\rho$-$\tau$-$y$ coordinate system
the contribution from the $\LL^{(3)}_0$
term may be expressed as:
\be \label{ewald}
\EE_0=8\, \pi \, \int \left. \, dy \, \sqrt{G_{yy}}
\,
{\p \LL^{(3)}_0\over \p \RR_{\rho\tau\rho\tau}}\, G_{\rho\rho}
\, G_{\tau\tau}\right|_{\rho=\rho_+}\, .
\ee
Here $\RR_{MNPQ}$ denotes the 
Riemann tensor computed using
the three dimensional metric $G_{MN}$ and
in computing $\p \LL^{(3)}_0/\p \RR_{MNPQ}$ we need to treat
$G_{MN}$ and $\RR_{MNPQ}$ as independent variables. 
In writing \refb{ewald}
we have used the fact that all terms involving covariant
derivatives of the Riemann tensor vanish in the BTZ black hole
solution.
Using the fact
that in
three dimensions $\RR_{MNPQ}$ can be expressed in terms of
$\RR_{MN}$ and $G_{MN}$,  and that for BTZ black hole
both $\p \LL^{(3)}_0/\p \RR_{MNPQ}$ and $\RR_{MNPQ}$
are proportional to $(G_{MP} G_{NQ}-G_{MQ} G_{NP})$,
\refb{ewald}
may be rewritten as\cite{9909061}:
\be \label{en0}
\EE_0 = \left. {4\pi\over 3} \, \, \left[
\int \, dy \, \sqrt{G_{yy}}
\,
G_{MN} \,
{\p \LL^{(3)}_0\over
\p \RR_{MN} }\right]\right|_{\rho=\rho_+}= \left.
{8\pi^2\over 3}
\, \rho_+ \, \left[G_{MN} \,
{\p \LL^{(3)}_0\over
\p \RR_{MN} }\right]\right|_{\rho=\rho_+}\, .
\ee
In order to evaluate the right hand side of \refb{en0} we note that
for the BTZ black hole solution given in eq.\refb{e16},
\be \label{en7}
\RR_{MN} = -2\, l^{-2}\, G_{MN}\, ,
\ee
and $\LL^{(3)}_0=h(l)$ according to eq.\refb{en2}.
Thus
\be \label{en8}
G_{MN} \,
{\p \LL^{(3)}_0\over
\p \RR_{MN} } = -{1\over 2}\, l^2\, \RR_{MN} \,
{\p \LL^{(3)}_0\over
\p \RR_{MN} } = 
-{1\over 2}\, l^2  l^{-2}{\p \over \p (l^{-2})} h(l)
= {l^3\over 4} h'(l) \, .  
\ee
Using \refb{en6} this can be written as
\be \label{en9}
G_{MN} \,
{\p \LL^{(3)}_0\over
\p \RR_{MN} } = -{3\over \pi l} \, g(l) +{1\over \pi} g'(l) 
= {3\, C\over 
l}\, ,
\ee
where in the second step we have used eqs.\refb{ecdef}
and \refb{en1}.
Hence \refb{en0} gives
\be \label{en10}
\EE_0 =  8 \, \pi^2\, C \, l^{-1}\, \rho_+ \, .
\ee

Let us now turn to the contribution $\EE_1$ from the
Chern-Simons term. For this we shall
view the black hole as a two dimensional solution and apply Wald's
formula. This gives
\be \label{en11}
\EE_1 = 8 \, \pi \, {\p \LL^{(2)}_1\over \p R_{\rho\tau\rho\tau}}
\, g_{\rho\rho} g_{\tau\tau}
= {4\, K\, \pi^2} \, {\varepsilon^{\mu\nu} F_{\mu\nu}
\over \sqrt{-\det g}} \, {1\over 2}\, 
g^{\rho\rho} g^{\tau\tau} g_{\rho\rho} g_{\tau\tau}
= - 8\pi^2 K l^{-1} \rho_-\, .
\ee
Thus the total entropy is
\be \label{en12}
\EE = \EE_0 + \EE_1 = 4\pi^2 l^{-1}\left( (C-K) (\rho_++\rho_-)
+ (C+K) (\rho_+ - \rho_-)\right) \, .
\ee

In order to express this as a function of
the physical mass $M$ and angular
momentum $J$ we need to relate $M$ and $J$ to
the parameters 
$l$ and $\rho_\pm$. We define
$M$ and $J$ as the conserved Noether charges
associated with time and $y$ translation symmetries respectively. 
The 
contribution
splits into a sum of two terms, one coming from the $\LL^{(3)}_0$
part and the other coming from the $\LL^{(3)}_1$ part.
The contribution from the $\LL^{(3)}_0$ part was evaluated in
\cite{9909061} and is given by,  
\be \label{em1}
M_0\pm J_0 = {2\, \pi\over 3\, l} (\rho_+\pm \rho_-)^2
G_{\mu\nu}  {\p \LL^{(3)}_0\over
\p R_{\mu\nu} } = {2\, \pi \, C\over l^2} \, (\rho_+\pm \rho_-)^2\, ,
\ee
where in the second step we have used \refb{en9}.
On the other hand
the contributions from $\LL^{(3)}_1$
were computed in \cite{0508218,0509148} and are given by:
\be \label{em4}
J_1 = -{2\pi K\over l^2} \, (\rho_+^2 + \rho_-^2)\, ,
\qquad
M_1 = -{4\pi K\over l^2} \, \rho_+ \, \rho_-\, .
\ee
Eqs.\refb{em1} and \refb{em4} now give
\be \label{em7}
M\pm J = (M_0+M_1)\pm (J_0+J_1) =  {2\pi (C\mp K)\over l^2}
(\rho_+\pm \rho_-)^2\, .
\ee
Substituting this into \refb{en12} and using 
the definitions of $c_L$, $c_R$ given in \refb{e26} we get
\be \label{e35}
\EE = 2\pi \sqrt{ c_L q_L\over 6} +  2\pi \sqrt{ c_R q_R\over 6}\, ,
\ee
where
\be \label{e34}
q_L={1\over 2} (M-J), \qquad q_R = {1\over 2} (M+J)\, .
\ee
This
is the correct expression for the entropy of a non-extremal black
hole in the presence of higher derivative and Chern-Simons
terms\cite{0506176}.

Note that the entropy and the near horizon geometry of 
the non-extremal BTZ 
black hole is determined by extremizing the function
\be \label{e33}
\EE = 2\pi \left[ q_L\, e_L + q_R e_R - \left({1\over e_L}
+{1\over
e_R}\right) g(l) + \pi \, K\left({1\over e_L}-{1\over e_R}\right)
\right]\, ,
\ee
with respect to $l$, $e_L$, $e_R$. 
Here
\be \label{ep1}
e_L= {l\over \rho_+ - \rho_-}, 
\qquad e_R = {l\over \rho_+ + \rho_-} 
\, .
\ee
We suspect that one can follow the procedure of \cite{0506177} to
manipulate Wald's formula and the equations of motion to derive this
extremization principle directly, but we have not so far succeeded in 
doing this.

\end{document}